# MERE RENOVATION IS TOO LITTLE TOO LATE:
## WE NEED TO RETHINK OUR UNDERGRADUATE CURRICULUM
## FROM THE GROUND UP

The last half-dozen years have seen *The American Statistician* publish well-argued and provocative calls to change our thinking about statistics and how we teach it, among them Brown and Kass (2009), Nolan and Temple-Lang (2010), and Legler *et al.* (2010). Within this past year, the ASA has issued a new and comprehensive set of guidelines for undergraduate programs (ASA 2014). Accepting (and applauding) all this as background, the current article argues the need to rethink our curriculum from the ground up, and offers five principles and two caveats intended to help us along the path toward a new synthesis. These principles and caveats rest on my sense of three parallel evolutions: the convergence of trends in the roles of mathematics, computation, and context within statistics education. These ongoing changes, together with the articles cited above and the seminal provocation by Leo Breiman (2001) call for a deep rethinking of what we teach to undergraduates. In particular, following Brown and Kass, we should put priority on two goals, to make "fundamental concepts accessible" and to "minimize prerequisites to research."

⎯⎯

George W. Cobb is Senior Research Professor and Robert L. Rooke Professor *emeritus*, Department of Mathematics and Statistics, Mount Holyoke College, South Hadley, MA 01075. (E-mail: *gcobb@mtholyoke.edu*).

The author is deeply grateful to David Moore for the shaping influence of his thinking and for the unique example of his expository style. My debt to David for both influences will be apparent to any reader who knows his work. I am grateful also to my many referees, whose suggestions and challenges led me to substantial rethinking and rewriting. I am convinced that the current version is much improved thanks to their thoughtful efforts, but if they disagree, the fault is entirely my own: I should not have paid attention to their comments.

## 1. BACKGROUND

Just twenty years ago, one of our ASA Presidents, Richard Scheaffer (1997) wrote "With regard to the content of an introductory statistics course, statisticians are in closer agreement today than at any time in my career." I agreed: I considered myself an enthusiastic part of that consensus. Today, however, I suggest that never in my professional lifetime has there been such a need to rethink our curriculum from the ground up, starting necessarily with alternatives to the former consensus introductory course, but with a more ambitious goal to rebuild the entire undergraduate statistics curriculum. In my 2005 address at USCOTS (U. S. Conference on Teaching Statistics) I argued that the standard introductory course, which puts the normal distribution at its center, had outlived the usefulness of its centrality (Cobb, 2007). This idea was in no way original with me. Its family tree goes back to the 1930s.[1] Despite its distinguished pedigree, the argument I presented at USCOTS was considered by some to be outside the mainstream, even radical. In the decade since then, I have come to regard my position in 2005 not as radical, but as far too conservative. Modern statistical practice is much broader than is recognized by our traditional curricular emphasis on probability-based inference. Our typical entry-level course follows the Advanced Placement syllabus (College Board 2010), with an emphasis on inference derived from the normal distribution, e.g., using the t-distribution for inference about two means or a regression slope. Even the newer, computer-intensive variants (e.g., Lock 2012 and Tintle 2015) still keep their emphasis on formal inference. Our typical upper division introduction for mathematics majors is an entire semester of probability followed by a semester of mathematical statistics.

In the last half-dozen years *The American Statistician* (*TAS*) has published a number of provocative articles about our subject and its teaching. Taken together, these articles prompt my conviction that we need to rethink our entire curriculum. Fortunately, Nicholas Horton and his colleagues have given us a well-researched and comprehensive kick-start in the form of a new set of Guidelines (ASA, 2014). These curricular guidelines (hereafter Horton report) recognize the seismic shift taking place beneath our feet: "The additional need to think with data – in the context of answering a statistical question – represents the most salient change since the prior guidelines were approved in 2000. Adding these data science topics to the curriculum necessitates developing …. capacities that complement more traditional mathematically oriented skills"(p. 7). These guidelines were appropriately constrained by a sense of what might realistically be expected in the near future. Realistic thinking has its virtues, but my premise is that long term there is also value to be found in more ambitious speculation. Some picnics beg for a skunk.

In this article I do not plan to address computer science in any detail. I can't possibly do better than refer readers to Nolan and Temple-Lang (2010, hereafter N&TL). Nor do I plan to write in detail about interdisciplinary research at the undergraduate level. Instead, I recommend the exemplary account in Legler *et al.* (2010, hereafter L+). A year before these two articles, Brown and Kass (2009, hereafter B&K) had written thoughtfully and provocatively under the title "What is Statistics? " I take these three articles, together with the Horton report as background.

In calling for a new curricular synthesis, I do not have a blueprint to offer.[2] I think it is far too early and the challenges are far too great for that. By my counting, it took from the mid-1950s to the mid-1990s to reach the last major consensus, and that one was limited to the introductory course. Not only is the scope of the current challenge broader (a four-year curriculum) but the impact of rapid advances in technology is harder to project forward. In what follows, Section 2 argues that our thinking about the undergraduate curriculum has become a tear-down, an aging structure that fails to take good advantage of the valuable territory on which it sits, and so imposes a steep opportunity cost on our profession and on our students. Section 3 argues that the three evolving roles of mathematics, computing, and context are a major source of the need for a new synthesis, and concludes with some issues to address, most especially a largely ignored tension between B&K and Breiman (2001). Section 4 summarizes Breiman's "Two Cultures" article, identifies some apparent conflicts with Brown and Kass, and offers some thoughts about reconciliation. Section 5 offers five guiding imperatives for thinking ahead about a new curricular synthesis, and Section 6 concludes with two caveats about implementation.

The outline I have just sketched tracks the argument I present here, but that outline leaves implicit my sense of three trends in the history of undergraduate statistics: the evolving roles of mathematics, computation, and context. A short version of my argument is that, taken together, these three parallel evolutions point toward one fundamental question adapted from B&K: How can our undergraduate curriculum be most effective in doing two things at once: making essential concepts accessible to intuition, and immersing students in early experience with authentic research? To do full justice to this question, we should start from scratch.

## 2. OUR THINKING ABOUT CURRICULUM HAS BECOME A "TEAR-DOWN"

I borrow my metaphor from the California real estate market, where territory has become so valuable that perfectly good structures once considered state-of-the-art and still acknowledged as serviceable have nevertheless been overtaken by rapid change, and risk losing out to more modern competition. For our profession, the valuable territory is the science of data; our competition takes place in the marketplace of ideas; and our statistics curriculum, though still serviceable, is increasingly at risk. "Big data" is one threat, from computer science; "analytics" is another, from business; "bioinformatics" is yet a third.

With reluctance, I have come to the conclusion that our consensus about curriculum needs to be rebuilt from the ground up. Our territory – thinking with and about data – is too valuable to allow old curricular structures to continue to sit contentedly on their aging assets while more vigorous neighbors take advantage of the latest ideas. Two of our most valuable assets are at grave risk: our profession's self-interest, and our profession's integrity.

### 2.1 Self-interest: The Danger to Our Field

"We don't get no respect." How often have we heard from colleagues some variant of that legitimate lament? Our canonical whine: Rarely has so much been accomplished by so few only to be ignored by so many, or, more formally "The field of statistics suffers from a lack of visibility and identity in spite of ever-increasing demands for statistical analysis ..." (ASA, 2008).

This challenge has been with us for ages, in a variety of forms. One old attack, from outsiders who put mathematical logic ahead of meaning-in-context, is that statistics is largely just a collection of recipes that can be learned by rote and applied without thought. A newer attack with more substance comes from those who put context first: learning enough mathematics to understand where our methods come from is an obstacle to simply getting the job done. For some, "the job" is no more meaningful than mechanically writing up a bibliography in the style required by your journal's editor: How many *s and what is the +/-? But from thoughtful scientists the job is to understand what the data have to say, and for too many of them, "we have been much less successful producing easy-to-master operating instructions and training programs" (B&K p. 105).

Either way, statistics suffers from the difficulty of its challenge to integrate abstract deductive thinking with interpretation in context. Older graduate-level books like Snedecor (1937) and Bliss (1967, 1970), outstanding books that relied on real data long before computing made real data the norm for introductory undergraduate courses, held little appeal for mathematicians. At the same time, elementary statistics courses in our client disciplines -- statistics in psychology, statistics for economics, business statistics – courses that often relied on context for meaning, competed for enrollments with courses offered in mathematics departments. (See Garfunkel and Young 1998.)

It is little wonder we felt then, and still feel, that others have been eating our lunch. The cliché of eating our lunch is not quite the right metaphor, however. It is more apt to say that the lunch we have been offering doesn't appeal to a broad enough clientele. We have insisted on seating only those who bring enough mathematics to the table, and who are in addition willing to sit patiently as we serve their meal linearly, one course at a time. Meanwhile, our competitors offer fast food. Their presentation may be inferior, and their diet may be heavy on the salt and fat of short term gratification, but customers can drive up, get the McNuggets of their virtual Happy Meal in a bag, and be done. On a loftier level, above the merely metabolic, there is a tension between the *ad hoc* pragmatism of context-based courses and the aesthetic unity of mathematically-based courses.

If our self-interest were the only issue at stake, I would not have my gorge up, but there is a much deeper reason to think hard about alternatives to McData's beckoning arches.

### 2.2 Integrity: We are honor-bound to preach what we practice

What we teach lags decades behind what we practice. Our curricular paradigm emphasizes formal inference from a frequentist orientation, based either on the central limit theorem at the entry level or, in the course for mathematics majors, on a small set of parametric probability models that lend themselves to closed-form solutions derived using calculus. The gap between our half-century-old curriculum and our contemporary statistical practice continues to widen.

Consider first probability-based inference. Although current practice now more often relies on randomization-based methods, few beginning courses make these ideas central (see Lock,*et al*. 2012 and Tintle *et al*. 2015 for just two of the few recent exceptions); current practice often relies on the bootstrap (see Lock *at al*. 2012 and Chihara and Hesterberg 2011 for two of the few contemporary books that make the bootstrap a central idea, and especially, see Hesterberg 2014); current practice increasingly relies on Bayesian as opposed to frequentist methods.

Even these more modern, computer-intensive methods for probability-based inference remain chained by mathematics to the paradigm of formal inference. The new curricular guidelines (ASA, 2014) are explicit about the importance of more flexible approaches to managing data and using data to solve problems that do not lend themselves to formal inference. For an easy example, consider the typical approach to regression. Regression models have many uses, most of which require no assumptions about probability models and use neither confidence intervals nor hypothesis testing. (See the list in Mosteller, Feinberg, and Rourke 1983 pp. 302 ff., and the book by Mosteller and Tukey 1977.) Nevertheless, almost all mainstream approaches to regression tie the model to the assumption of independent normal errors and the consequent inferences. (See Cobb, 2011 for more detail, and an alternative approach to regression that includes formal inference but puts it last, toward the end of the course). Teaching of regression methods, even if inference is postponed until late, nevertheless belongs to the mainstream. A question that leads us in a different direction: We teach paper-and-pencil EDA (exploratory data analysis) in a first course; why not teach CEDA (computer-aided exploration, or algorithmic data analysis) as well? Classification and regression trees (see Breiman *et al*. 1984) can be made accessible at the level of a first course. In an innovative approach that explores the frontiers, Amy Wagaman (2013) at Amherst College has developed an entry-level course that introduces multivariate methods to students with no previous background in statistics and no mathematics beyond high school algebra. We can all learn from her example.

I want to be clear: Although I have called for rethinking the curriculum from the ground up, I am not advocating a scorched earth approach. I am convinced that there is much of value in our standard curriculum.[3] All the same, our territory – thinking with and about data – has become too valuable, and the options for new topics and courses have become too many and too compelling, for us to continue to rely solely on riding the aging war horses of our past. Pegasus beckons.

To be concrete, consider, for example, a project used by N&TL in their computing course (p. 104): "The data consist of … over 9000 e-mails … as raw text files … Students write functions to extract … dozens of variables … fit statistical models to predict spam … using recursive partitioning / classification trees, and … assess how well their method works on test data …"

Now, as a chastening contrast, imagine a three-part triage: (1) Does your data come from a single source? If no, go away. (2) Does your data fit the standard cases-by-variables format? If no, go away. (3) Can you justify using a probability model? If no, go away. Such an attitude narrows our clientele, but despite the harm to our profession, that bottleneck is where we have gotten stuck in our teaching. Bottom line: We must get un-stuck. The next section hypothesizes about how needless dependence on mathematics has made our thinking sticky, and how computing can help us open up new possibilities.

## 3  MATHEMATICS, COMPUTING, AND CONTEXT:  THREE CONVERGING TRENDS

There are many threads that trace paths through the history of statistics and its teaching. This section follows three: the roles of mathematics, computation, and context. My thesis is that these three show converging trends that make our undergraduate curriculum both victim and beneficiary. Possible paths forward will be left implicit here but will be addressed in Sections 5 and 6.

### 3.1  How we got stuck:  The evolving role of mathematics

The changing role of mathematics leads to useful thinking about where statistics and its teaching have come from and where we may be headed. In an admitted oversimplification, I reduce the whole enterprise to a succession of four stages: Bernoulli, Fisher, Mosteller, and Advanced Placement (AP).

1. *Bernoulli:  Mathematics as computational engine.*  As I see it, the first use of mathematics to address a truly deep statistical question is due to Bernoulli, who wanted to quantify the relationship between sample size and margin of error. From Bernoulli in 1692 through the next two centuries and more, mathematics served statistics mainly as a computational engine.[4]

2. *Fisher:  Mathematics as source of unifying theory.*  Fast forward 230 years from Bernoulli to Fisher's 1922 paper in which he maximized likelihood to derive estimators, factored the likelihood to define sufficient statistics and ancillary statistics, and used the variance of the log derivative of the likelihood to define efficiency. Fiducial inference is based on an assumed symmetry between pre-data $x$ and post-data $\theta$ in the likelihood function.[5]

3. *Mosteller:  Mathematics as a source of respectability.*  In the mid-1950s, Frederick Mosteller (and others) used probability as a way to insert the camel's nose of data analysis into the tent of the undergraduate curriculum. Ever since, what eventually was to become our Stat 101 course has relied on mathematics as a way to justify its attention to real data.[6]

4. *AP statistics: Mathematics as obligatory presence.* For the last half-century, mathematics has been hovering like a helicopter parent, making it hard for our Stat 101 course to go out and play with its curricular friends. In our current day-care center the central limit theorem is bully. The goal is to get students to tests and intervals based on the normal approximation. Moreover, the object of those inferences is always the center of some probability distribution, either the mean or a binomial proportion. We teach inference first for one mean and one proportion, then for the difference of two means and of two proportions, and then, if there's time, we teach inference for several means or proportions, and for the conditional mean of Y given x, provided that relationship is conveniently linear.

Our traditional statistics curriculum relies heavily on its connections to mathematics. At all levels, from introductory to advanced, the mathematical content is a deterrent to some students who could learn the important statistical content in some other way. This barrier also keeps some applied scientists from learning the statistics they need, and makes non-statistical thinking an attractive alternative.

Thanks to computers statistics no longer needs mathematics as a computational engine. Thanks to the ubiquitous media attention to "big data" statistics no longer needs mathematics as a source of respectability. What then can mathematics offer us to make the effort worth it? My answer is that although we need continuing vigilance to ensure that mathematics is not an obstacle, mathematics remains essential. Mathematical thinking has served, since Plato, as the purest model for one of our most direct paths to deep understanding of patterns and connections, namely, abstraction-as-process. (This role for mathematics in our thinking about curriculum is illustrated Sections 5.2 and 5.4a.)

Over the last 70 years the evolving role of mathematics in statistical practice and in our teaching of statistics has been driven and shaped by the evolving role of computers.

### 3.2 Computers: Revolution? (No) Reformation? (Yes)

The phrase "computer revolution" has become a cliché, and we still hear it often, as a metaphor not so much dead as deadly, in the sense of killing thought. The Metaphysical Poets gave us a richer, more vital kind of metaphor, the conceit, an extended metaphor with explicit correspondences. In this section I suggest that it is useful to think about the effect of computers through several explicit parallels with the Reformation. Although it would be easy for readers to dismiss these parallels as contrived, I hope that won't happen, because I think the parallels are strong and deep, and taken together, they offer a useful mnemonic for keeping in mind an important cluster of ideas. By way of introduction, here are four developmental stages related to computers.

- *Single step.* In the early years, computers allowed us to do messy arithmetic quickly and cheaply. This led us to teach with real data, to fit multiple models, and to expand our use of graphics and diagnostics. Here we were relying mainly on those components of analysis, like fitting a single model, that could be completed in just *one step.*

- *Several steps*.  Computing also made it possible to make routine use of existing iterative methods like logistic regression, and later, generalized linear models.  The EM algorithm brought an abstract unity to a host of methods for incomplete data (Dempster *et al.* 1977).  These iterative methods took *several steps* to get sufficiently close to a solution, but "several" could typically be counted with one or at most two hands.
- *Thousands of steps*. More ambitiously, computers allowed routine use of methods that came to be known as "computer intensive" – methods like randomization-based inference and the bootstrap, that required thousands or even tens of *thousands of steps*.  (See Diaconis and Efron 1983.)
- *Bayes*.  More recently (see Gelfand and Smith 1990) Markov Chain Monte Carlo and related methods such as multiple imputation (Rubin 1996) have led to widespread use of Bayesian methods for applied work, which use, in turn, has led to a major reversal of an earlier prejudice against what had long been dismissed as an inappropriately subjective approach to data analysis.  Decades before Gelfand and Smith, Savage (1954), Birnbaum (1962), De Finetti (1972) and others had argued rigorously that if you were not a Bayesian, you were incoherent.  Statisticians read the arguments, followed the proofs, nodded in agreement, and continued in their pursuit of incoherence.  *It was the computer, not logic, that persuaded our profession to embrace Bayes.*  In statistics, practice usually leads and theory follows, rarely the reverse.

My brief chronological sketch is meant to show how the impact of computing on our thinking has changed as capacity has grown.  What began as "we can do the same things as before, only faster and more easily" grew to spurring the development of new methods (generalized linear models, the bootstrap, EM) and eventually to a major shift in orientation away from a long-standing reluctance to use Bayesian methods.

In short, few statisticians now think of the computer as merely bringing us a faster way to do the same old things.  I suggest that something similar to the invention of the computer has happened only once before in the last thousand years of our history:  the invention of the printing press. Initially, it would have been easy to think of the printing press as "merely bringing us a faster way to do the same old things," in this instance a faster way to make copies of manuscripts.  In hindsight, of course, we recognize that Gutenberg's way of "doing things faster" not only led to wider distribution of the Latin Bible, but also inspired multiple translations into the vernacular, which led in turn to diminishing the role of priests as guardians of orthodoxy, and eventually to the emergence of Protestant sects.[7]

I see the same sort of thing as once happened with the printing press now happening with computing, not just in statistics, but in communication generally via the web.[8]  Much as learning Latin was once a challenging prerequisite to reading the Bible, in statistics facility with mathematics has been a prerequisite to understanding and using methods of data analysis.   The select few who knew enough mathematics were a kind of priesthood.  Just as movable type inspired translations that bypassed the barrier of Latin, computer software and computer-intensive methods have made statistical methods broadly available to those who are not mathematically facile, and unfamiliar with probability.  Big data, bioinformatics, and analytics – varieties of computer-aided thinking -- are our heresies.   They rely on computers to circumvent the need for mathematics.

Whereas the roles of mathematics and computation have been evolving along paths that have not turned back on themselves, the role of context is different.

### 3.3 Context:  A Return to Our Roots

David Moore put it succinctly:  "Data are numbers, but they are not 'just numbers.' They are numbers with a context" (Moore and Notz 2006 p. xxi).  That short summary captures a core idea that in the past has all too often eluded teachers of the introductory course.[9]  Fortunately those days of yore are largely behind us, but looking to the future, one can see in B&K and L+ a knightly gauntlet thrown at our feet.  Before addressing their challenge, however, I find it helpful to review the changing role of context over the last hundred years.

The earliest uses of examples in books for teaching statistics date back to the first decades of the twentieth century, in books intended for graduate students and research workers, often in agriculture.[10]  These were followed by other, less technically demanding books, also aimed at particular areas of application, rather than for general undergraduate audiences,[11] but they can arguably be seen as helping to bridge the gap between books with a particular emphasis aimed at professionals, like Snedecor (1937) and books with an emphasis on probability aimed at undergraduates, like Mosteller (1961).  In between, and also helping to bridge the gap, starting in the 1950s, a number of books at the level of popular science appeared. [12]  These books used the stories of real examples for exposition.  By the 1970s, introductory statistics books intended for a general college audience began to appear, and by the late 1970s some were using real data to illustrate statistical theory and methods.[13]

At roughly the same time, we had other books that were intended not as main textbooks, but rather as supplements and complements to be used in conjunction with some other book.  In these books the applied examples served as a primary vehicle for exposition.[14]  More recently, from the 1990s forward, we have seen a number of textbooks for project-based courses.[15]  In this chronology I see a progression toward a greater and greater role for applied context, a progression which I and many others see as a challenge for our curriculum:  How far *can* we go, and how far *should* we go in our reliance on context for teaching statistics?

My thesis, borrowed from B&K and bolstered by examples cited in Section 5, is that as a profession we have only begun to explore the possibilities.  The history of our subject also supports this thesis:  Unlike probability, a scion of mathematics, statistics sprouted *de novo* from the soil of science.  Its roots run back to astronomy and geodesy[16] (Stigler, 1986).  Over time, as statistics came to be recognized as a subject in its own right, the quest for abstract understanding, as sketched in 3.1 above, often put applications in the secondary role of mere illustrations.  We now seem poised for a Renaissance of context, a rebirth of our reliance on research as integral to learning our subject.

Before exploring some possibilities Section 4 first reviews and expands on another challenge to our profession, Breiman's (2001) declaration of two statistical cultures.

### 4   LEO BREIMAN'S TWO CULTURES AND A SECOND LOOK

In a seminal article fourteen years ago, Leo Breiman (2001) presented a forceful argument that statistics is divided by fissures that separate academics from practitioners, and more importantly, separate two ways of thinking, which he called the "stochastic" and "algorithmic" cultures. Eight years later B&K offered a different view. I don't read their article as meant to challenge Brieman in any direct sense, and I suggest the two articles are complements that raise important questions when taken together. This section first summarizes Brieman's two cultures (4.1), then offers a second look (4.2) in light of Brown and Kass (2009), and ends with some attempts to resolve apparent conflicts.

### 4.1 Breiman's Two cultures: "stochastic" and "algorithmic"

The "stochastic" culture (98% of academic statisticians, according to Breiman) starts from a probability model, and proceeds deductively to estimators, test statistics, and their distributional properties. "Nature" is regarded as a black box that converts a collection **x** of input values to a collection **y** of output values. The stochastic modelers try to find a workable probability model for what happens inside Nature's black box. Patterns relating **x** and **y** can provide information about the model, and the model can be used to evaluate particular methods, e.g., to derive properties of estimators or test statistics. The algorithmic culture is more direct, bypassing Nature's black box in order to focus directly on the relationship between **x** and **y**. In the spirit of my Reformation metaphor, the stochastic culture is the orthodoxy of the Medieval Roman church, the algorithmic culture is the technologically enabled Protestant heresy.

The traditional normal-based two-sample *t*-test is a canonical instance of the stochastic approach to modeling. A classification tree is a canonical instance of the algorithmic approach to modeling. (See Breiman *et al.* 1984. For an example used in teaching, see N&TL p. 104.) Four features of the examples deserve attention. (1) The traditional t-test relies on comparatively strong assumptions about the observed values; the classification tree does not. Thus the algorithmic method tends to be applicable to data sets that do not satisfy the distributional requirements of the t-test.[17] (2) The t-test starts from an assumed probability model; the classification tree starts with the goal of the analysis. (3) The justification for the t-test is abstract, based on deductive properties of the model; the justification for the regression tree is empirical, based on the misclassification rate observed directly from a test sample. (4) The t-test is not intuitive and is correspondingly hard to explain; the classification tree is comparatively much simpler.[18]

I suggest that a useful way to understand Breiman's dichotomy in the context of our statistics curriculum is though comparison with Tukey's (1977) *Exploratory Data Analysis* (EDA). EDA was novel in two deliberate ways: (1) It did *not* rely on a probability model, and (2) it did *not* rely on technology. The first negative, no probability, is preserved in our curriculum. Most introductory courses now begin with some version of EDA *before* any probability. Randomized data production comes after, and only then do students meet probability models for data. Courses beyond the first one take EDA for granted and use it freely. The second negative, no technology, is now largely forgotten, because we tend to take technology for granted, but back in the late 1960s, calculators were so heavy and so expensive that few undergraduates ever got close to one. Time-shared terminals were in their infancy, and laptops were decades in the future. There was a gigantic void between "introductory statistics" and "real data." Tukey's second innovation bridged that gap. His stemplots and five-number summaries made it possible for students in a first statistics course to work with real data to investigate real questions using pencil and paper only. Beyond counting no arithmetic was needed.

Tukey's first innovation, exploratory methods that did not rely on probability, introduced a new approach to data, but was constrained at the time by limited access to technology.[19]  Looking back, a natural question arises:  What if Tukey were alive today?  What would contemporary EDA look like if it were to take full advantage of technology in order to seek patterns in large data sets?  For me, one answer is the approach to data that Breiman called "algorithmic."  Table 1 offers a comparison:

| | Probability Model | Intuitive/ Exploratory | Technology | Cross-validation | Data Size | Defined goal |
|---|---|---|---|---|---|---|
| Tukey | No | Yes | No | No | Small | No |
| Breiman | No | Yes | Yes | Sometimes | Large | Sometimes |

Table 1.  Tukey and Breiman compared

In short, we might think of Breiman's "algorithmic thinking" as a computer-aided extension of Tukey's EDA, or CEDA.[20]  What unites Tukey's EDA and Breiman's CEDA is their reliance on empirical exploration and their independence from any probability models for the data.   That reliance and independence may seem at odds with B&K.

## 4.2  Brown and Kass:  "What is statistics?"

Eight years after Breiman, B&K, in answering their title question, "What is statistics?" began their definition with "Statistics uses probabilistic descriptions of variability … " (p.107).  Their characterization that statistical thinking is necessarily based on a probability model would seem to exclude Brieman's "algorithmic" culture as non-statistical, and in fact B&K state that "our formulation cannot accommodate the perspective of Breiman" (p. 110).[21]  They strike me as careful not to issue a direct challenge to Breiman's thinking, and I suggest that the polite tectonic collision between the two points of view pushes up a ridge rich with the ore of issues that a new synthesis must address:

- Do we really want to cede to others all methods of data analysis that do not rely on a probability model?  (WWTS:  "What would Tukey say?")
- Probability is a notoriously slippery concept.  The gap between intuition and formal treatment may be wider than in any other branch of applied mathematics.[22]  If we insist that statistical thinking must necessarily be based on a probability model, how do we reconcile that requirement with goals of making central ideas "simple and approachable" (B&K p. 108) and minimizing "prerequisites to research" (B&K p. 108)?
- Is a probability model the only way to assess the credibility of an analysis?  What about cross-validation for a data set that is "just there" as in N&TL p. 104?  Without a sampling model, there are dangers about scope of inference, but we can teach these dangers without a formal probability model.
- Breiman located the probability model in the middle box ("Nature") between input **x** and output **y**.

- o Is a stochastic model for what goes on in that box required for thinking to be statistical?
- o Can the idea of variability be used to assess methods that qualify as "algorithmic" in Breiman's sense?

These questions are meant as a challenge, but not a challenge to the authors B&K. We owe them thanks for bringing these issues to the surface. The challenge is to our profession, and to all of us who seek to help shape the undergraduate curriculum. Breiman's distinction between his two cultures splits at the role of probability models, present or absent, but in a different way. B&K's answer to "What is Statistics" also splits at the role of probability, present or absent. I find these distinctions both important and useful. I also consider them more fruitful than sharp. For me, the fuzz at the supposed cusp is an invitation to follow their lead.

Tukey's boxplots are quantiles for the eye. Are quantiles probability-based? (I assume yes. Otherwise they would not qualify as "statistics" in the sense of B&K.) Breiman uses misclassification percentages to evaluate classification trees. Does that make trees "stochastic?" (I assume no, because there is no probability model from which the tree is derived.) The EM algorithm and Fisher's method of scoring are certainly algorithmic in an old sense, but they maximize likelihood and so are "stochastic" but not "algorithmic" in Breiman's sense. In asking these questions, I am not seeking answers so much as trying to reinforce the point that even we statisticians -- we who are trained to be modest about how little we know -- are now at a point where, when it comes to the undergraduate curriculum, we know much less than we thought we did twenty years ago. For me, the lesson is clear, that we should experiment with boldness but conclude with diffidence.

Meanwhile, I think it is useful to distinguish between Breiman's "stochastic" use of probability models for Nature's black box, and B&K's "statistical" use of probability to quantify the efficacy of methods. I also think it is useful to distinguish between Breiman's use of "algorithmic" to describe computer-aided exploratory methods (CEDA) and the older use of "algorithmic" to describe numerical approximations to exact solutions. The next section, on principles for thinking about curriculum, relies on these distinctions.

## 5.  LOOKING AHEAD:  FIVE IMPERATIVES

I hope the previous sections have been persuasive about the need to rethink the undergraduate statistics curriculum. We need to rethink the role of mathematics, rethink the role of computing, and rethink the role of context. These three bundles of roots run so deep, and are so intertwined, that just grafting a new branch or two onto the old stock will not continue to bear fruit over the long term. In my opinion, the Horton report is a critical step forward but we are far from ready to settle on a new curriculum. The challenges and opportunities are still taking shape. We need to experiment aggressively. In the hope of moving us forward without suggesting a premature consensus, this penultimate section offers five guiding imperatives, or, to revive and stretch my Reformation metaphor, five theses. Lest you think I am offering myself as a Martin Luther, note that my paltry five leave me 90 short. My shortcomings aside, we still need a Reformation.

If we are truly to rethink our curriculum at a deep level, we ought to start with foundations. By foundations I do not mean which concepts and content. I am convinced we will need an extended period of ferment, experimentation, and settling out to reach a new consensus on content, much as it took us decades to reach the old consensus on the now middle-aged introductory course.  In this section, I try to remain agnostic about content in order to focus on five broad imperatives that I would like to see guide our efforts.  All five expand from B&K's own imperative, "Minimize prerequisites to research," which, I argue, can be broadened, made more ambitious, and applied across our curriculum.  B&K decry what they call "first understand, then do" (p. 108).  I agree, but I worry that their "Minimize prerequisites to research" may be misconstrued to suggest that research is the dessert, prerequisites are the spinach.  In what follows, I divide and broaden their directive:  Flatten prerequisites *and* teach through research.

Because Brown and Kass teach at major research universities and have published distinguished articles in neuroscience, it would be natural for readers to interpret "research" to mean publishable, professional-level research of the sort the authors have done.[23]  I argue (1) that "research" should be understood more broadly, namely, using data to study an unanswered real-world question that matters, and (2) that teaching through research has been proven successful at all levels, *beginning as early as elementary school*.

My five imperatives, with a debt to Brown and Kass, are 5.1 Flatten prerequisites, 5.2 Seek depth, 5.3 Exploit context, 5.4 Embrace computation, and 5.5 Teach through research.  In the abstract these may seem hard to argue against, like motherhood and apple pie, but in this section I try to be concrete enough through examples to provoke some opposition.   Three recurring examples are (1) design and analysis of experiments (ANOVA), (2) Bayesian thinking via Markov Chain Monte Carlo, and (3) a variant of the traditional mathematical statistics course.  I have taught these multiple times as first statistics courses at a comparatively elementary level.  My point is not that we *should* teach these as first statistics courses, but rather, precisely because we *can*, to assume that we can't limits our thinking.

### 5.1  Flatten prerequisites

In our practice of applied statistics, we ourselves routinely flatten prerequisites, using a "just-in-time" approach to what we need to know (Cobb 1993, Rossman and Chance 2006 preface).  For example:  How many statisticians who have made contributions to the statistical analysis of data from microarrays have taken the prescribed sequence of courses in introductory biology, introductory chemistry, organic chemistry I and II, molecular biology, genetics, etc.?  In our profession as we practice it, we often wait to learn what we need to know until we need to know it, and we focus our learning on what we need to know.  Why shouldn't the ways we teach our subject follow the approach we use in practice?  Here, yet again, I agree with B&K that what we do in practice is at odds with how we teach, and I suggest that, yet again also, what we teach is shaped by a mathematical aesthetic – to build a glorious structure – rather than shaped by a pragmatic approach to solving real problems. [24] Here, with some grateful borrowing from mathematicians O'Shea and Pollatsek (1977), are some of the concerns about prerequisites, along with my sense of how they bear on our curriculum in statistics:

- *Prerequisites limit enrollments.* For example, the traditional approach to mathematical statistics puts the course at the end of a five-course sequence:
    Calc I → Calc II → Calc III → Probability → Math Stat[25]
    Arguably this one structure does more to block our profession's pipeline than any other aspect of our undergraduate curriculum.  Sections 5.2 ff.  offer an alternative approach that requires only one semester of calculus.

- *Prerequisites limit course choices for students.* Currently and typically, the introductory applied course is the needle's eye through which all students must pass to get to other applied courses. As the examples will illustrate, this prerequisite is not necessary.
- *Prerequisites limit faculty in the courses they can offer.* If your teaching load is six courses and your prerequisite structure requires you to teach four sections of Stat 101, wouldn't you prefer more variety?
- *Flattening prerequisites encourages faculty to think about which skills and concepts can be taught as needed instead of requiring an entire course in advance.* Example: Partial derivatives are intuitive geometric concepts, as are multiple integrals.[26] Technical skill is not needed for fundamental concepts.
- *Flattening prerequisites encourages creative thinking about new courses.* Danny Kaplan (2009) at Macalester College has created a beginning course that not only includes most of the usual topics, but also relies on the geometry of Euclidean *n*-space to help students think about causality, confounding and adjusting in the context of multiple regression.[27]

Flattening prerequisites leads naturally to a broader set of options and a more flexible course structure. Of course there are costs to that more flexible course structure. Creating a new course takes more time and effort than teaching an established course that comes with supporting materials. Teaching students with a broader range of backgrounds is harder than teaching a more uniform group. All the same, I have concluded that on balance, our current prerequisite structure limits our students' choices, constrains our faculty, and hurts our profession. Here, as a deliberate provocation, are three questions: (1) How would you teach a student with no background in statistics to design and analyze a multifactor experiment with both fixed and random effects, both crossed and nested factors? (2) How would you teach Markov Chain Monte Carlo and Bayesian hierarchical models to a student with no background in statistics, calculus, or probability? (3) Can we rethink our introduction to the concepts of statistical theory so as to offer it to first-year students?

### 5.2  Seek Depth

By seeking depth I mean stripping away what is technical – formalism and formulas – in order to reveal what is fundamental, in order to "represent the discipline as being rich in profound concepts" (B&K p. 108). I part company with B&K, however, when they write "central ideas must *remain* simple and approachable (p. 108, my italics). At the undergraduate level I suggest that central ideas must *be made* accessible and approachable. As a goal, we should seek a way to summarize profound concepts simply and succinctly, in words only. That challenge requires a deep understanding, and that is where mathematics is essential. Here are two examples. As you read through each one, you might challenge yourself to create a short description for students of the central idea.

*Analysis of Variance (ANOVA) and symmetry.*

What are the essential ideas here? Most traditional textbooks take one of three over-narrow points of view: ANOVA generalizes the two-sample t-test to three or more samples,[28] ANOVA is a structured comparison of variance estimates via messy algebraic methods swathed in mysterious bundles of subscripts,[29] or ANOVA is a special case of regression whose predictors are indicator variables.[30]

What all these myopic approaches miss is the fundamental connection between designed experiments and planned groups of symmetries in the data. *The basic idea is that if your experiment is well designed, you can take apart your data into independent pieces that correspond to the influences you care about, and additional pieces that correspond to influences that would otherwise get in the way.* The enabling mathematics is advanced, a form of abstract harmonic analysis applied to finite groups: planned groups of symmetries determine invariant subspaces in an *n*-dimensional Euclidean space, and projection of the response vector onto these subspaces determines the ANOVA. (See Fortini 1977 and Diaconis 1988.) Students do not need to know any of this. For balanced designs the least squares projections onto orthogonal subspaces can be computed *and explained* at an elementary level as differences of group averages (estimated effects) determined by a single, simple algorithm that not only gives a breakdown of observed values into components, but also gives the parallel decompositions of sums of squares and degrees of freedom (Section 5.3), and, in addition, gives the expected mean squares and F-tests. As far back as the 1950s some textbooks (e.g., Fraser1958) presented one-way and two-way ANOVA as a decompositions of the data into tables of estimated effects. More recently, at a more elementary level, so did Mosteller *et al.* (1983). Hoaglin *et al.* (1991) is in the same spirit, and evocatively names the tables of estimated effects "overlays." By relying on a single, simple algorithm (Cobb, 1984), I have long taught ANOVA and design (as far as multi-factor designs with both crossed and nested factors, both fixed and random effects) as a *first* statistics course with *no* prerequisites other than high school algebra II. (See Cobb, 1998, Preface.) Moreover, a focus on planned symmetries for analysis makes it comparatively easy to spend much more time actually *designing* experiments as opposed to merely analyzing those designed by someone else.

*Conditional probability and Bayesian inference*

Here, as in the previous ANOVA example, the usual emphasis on mathematical formality and technical details gets in the way of teaching the core concepts. The swaddling in this instance is built into a definition: $P(A|B) = P(A \cap B)/P(B)$. Conditional probabilities are introduced *after* "ordinary" probabilities, as a separate kind of entity, by way of the defining fraction. What often gets lost is that in practice *all probabilities are conditional* (conditional on the choice of sample space $\Omega$), and that given the sample space, all discrete probabilities reduce to a fraction (or limit) equivalent to #$A$/#$\Omega$. For $P(A|B)$, we simply restrict the sample space to outcomes that satisfy the condition $B$, then compute an "ordinary" probability. The only useful thing to come from $P(B)$ in the denominator is to remind us that probabilities should be normalized to add to 1.[31]

A major advantage to computing P(A|B) by restricting the sample space is that it removes one of the major impediments to teaching Bayesian inference at an elementary level, namely, the reliance on multivariable calculus to compute denominators (marginal probabilities) of the form $P(B)$. In most genuine applications of Bayesian inference, the marginal probability in the denominator is a multiple integral. This denominator, a computational dog's breakfast, has nothing to do with the fundamental concept, and our

insistence that we regard Bayesian inference as an advanced topic is tantamount to letting the breakfast eat the dog.

*The basic idea, which needs no calculus, goes back to Laplace and to what I have called his "data duplication principle," viz., "a parameter value is believable to the degree to which it reproduces the observed data", or, in modern terms, the posterior is proportional to the likelihood.*[32] Our profession's obeisance to technical mathematics has misled us into thinking that we can only teach Bayesian thinking to students who can do those multidimensional integrals in the denominator.

However, if we take $\#A/\#\Omega$ as the fundamental concept and introduce more complicated probabilities via approximations of the same form, we open up a simple path to Bayesian methods via simulation as described in Section 5.3. I have used this approach for many years in teaching Bayesian inference via MCMC in a class for students with no previous coursework in either probability or statistics, and no calculus. This assertion may suggest that the course must be superficial, a kind "statistics appreciation," but such an inference is part of our mindset. In the discrete world of simulation, all sample spaces are finite and all probabilities are fractions. "Discrete" and "deep" are not mutually exclusive.

*Concepts of Statistical Theory (Mathematical statistics)*

As a thought experiment, run through the basic concepts and theory of estimation. Note how almost all of them can be explained and illustrated using only first-semester calculus, with probability introduced along the way. For example, a focus on discrete distributions with continuous parameters means that moments are sums, not integrals, but the likelihood is differentiable, so all the associated theory is there to be developed. Students can create estimators evaluate their properties (bias, variance, consistency), and compare methods of estimation (moments, least squares, maximum likelihood). The centrality of the score function in Fisher's theory is accessible also, e.g., it's use for maximizing likelihood, its moments, information and efficiency, and the Cramer-Rao inequality with its elegant one-line proof.[33] Treating the normal as a family of continuous approximations to discrete distributions fit by matching moments opens the way to asymptotic tests and intervals.

If we are truly to "minimize prerequisites to research" (B&K p. 108) we can't continue to treat subjects like ANOVA, applied Bayes, and mathematical statistics as advanced or even intermediate-level topics. If we can teach them at a deep level with no prerequisites, surely students can learn them as part of a research project. Deep understanding can help us isolate the essentials and find an elementary way to explain them. More often than our curriculum reflects, we can do this by relying on computation.

### 5.3 Embrace computation.

Section 4.2 distinguished two quite different meanings of "algorithmic." First, following Breiman, we can think of computer-based exploration as an important part of statistics that we should teach more often and earlier in the curriculum as recommended in the Horton report. We statisticians do not even try to teach algorithmic approaches at the elementary level. Computer scientists do, but we don't. In urging that we rethink from scratch what we teach I am hoping that we do more than just insert a single new "big data" unit into an existing course, more also than just insert a new computing course into the existing curriculum.  To seek effective ways to develop a curriculum that teaches both of Breiman's two approaches will be a major change for us, and I suggest that it is far from clear where we will end up. Getting there, wherever "there" turns out to be, will require a lot of experimentation with new courses, and a lot of time and thought, before we reach the understanding we will need for a consensus curriculum.  One such experiment is Amy Wagaman's (2013) first-year course in multivariate data analysis at Amherst College.

　　We can also embrace the older sense of "algorithmic" as referring to numerical methods for finding approximations to such things as integrals and derivatives.  Here too, we have overlooked a major opportunity.

*Design and ANOVA*

　　In many ANOVA courses with an applied emphasis the traditional algebraic notation is a major barrier.  Consider, for example, the formulas for sums of squares with their multiple subscripts, bars, dots, and summation signs.  Each new design brings a new set of such formulas, any attention to indices distracts from ideas, in particular, from the fact that the same single algorithm applies to all balanced designs.  Relying on the computer output and ignoring the formulas is in many ways an improvement, but the path from the data to the ANOVA table remains something of a black box.  As an alternative, consider (Figure 1) the  decomposition into overlays of data from a complete block design with four subjects and three treatments.

|        | A | B | C |   | Grand Ave. | | | | Block effect | | | | Treatment | | | | Residual | | |
|--------|---|---|---|---|----|----|----|---|----|----|----|---|----|----|----|---|----|----|----|
| Subj 1 | 13 | 6 | 5 |   | 10 | 10 | 10 | | -2 | -2 | -2 | | 4 | -1 | -3 | | 1 | -1 | 0 |
| Subj 2 | 19 | 11 | 9 | = | 10 | 10 | 10 | + | 3 | 3 | 3 | + | 4 | -1 | -3 | + | 2 | -1 | -1 |
| Subj 3 | 13 | 9 | 8 |   | 10 | 10 | 10 | | 0 | 0 | 0 | | 4 | -1 | -3 | | -1 | 0 | 1 |
| Subj 4 | 11 | 10 | 6 |   | 10 | 10 | 10 | | -1 | -1 | -1 | | 4 | -1 | -3 | | -2 | 2 | 0 |
| SS | 1364 | | | = | 1200 | | | + | 42 | | | + | 104 | | | + | 18 | | |
| df | 12 | | | = | 1 | | | + | 3 | | | + | 2 | | | + | 6 | | |

Figure 1.  Decomposition into overlays of data from a complete block design
The overlays show the decomposition of the observed values into pieces.  (This is the connection to harmonic analysis.)  Factors are represented visually, as partitions of the data into groups that corresponds to meaningful groupings, in this case the grand mean, the blocks, the treatments, and residuals.  The algorithm has three parts:  (1) The overlay for a factor is computed as its set of group averages minus the overlays for all "outside" factors.[34]  (2) To get the sum of squares for a factor you simply square and add the numbers in its overlay.  (3) The degrees of freedom equals the number of "free" (distinct) numbers not fixed by adding to zero.

　　Of course this compact summary is too much too fast for a beginning student, but the one-way design gives an easy start and each new design repeats the same pattern, so that abstract understanding emerges through practice.  Moreover, the overlays correspond to meaningful features of the design and of the data, so there is nothing mysterious about the algorithm.  The connection to computer output is direct and intuitive, and the focus on meaningful partitions reinforces the connection between the design and the analysis.

*Bayes and MCMC*

The key idea, set out in Section 4.2, is Laplace's "data duplication principle," that a parameter value is believable to the degree that it reproduces the observed data." This principle is easily turned into an algorithm:  Given observed data $y_{obs}$,

1. *Generate* a sequence of parameter values $\theta$ according to the prior.
2. For each $\theta$ value, *simulate* a response value $y_{sim}$.
3. *Compare*:  If $y_{sim} = y_{obs}$, keep the value of $\theta$, which has successfully reproduced the observed data; otherwise discard that value.
4. *Estimate*:  The saved values of $\theta$ follow the posterior.

To be clear, this is definitely *not* an efficient algorithm for computation.  Its purpose is to illustrate for students the reasoning behind Laplace's approach to Bayesian inference.[35] Once students understand the principle, they are ready to learn more efficient algorithms such as Metropolis-Hastings and the Gibbs sampler.

*Mathematical statistics:  concepts versus calculus*

Our junior-senior course in "mathematical" statistics sits serenely atop its mountain of prerequisites, unperturbed by first and second year students.  Tradition takes it for granted that they must climb through the required courses in calculus and probability before ascending to maximum likelihood and the richness of profound ideas. Of course we *want* students to learn calculus and probability, but it would be nice if we could join all the other sciences in teaching the fundamental concepts of our subject to first year students. Here, as in other areas of our curriculum, what we teach lags behind practice.  Statisticians rely on simulation to estimate probabilities and moments that are beyond the reach of their analytic skills; we should teach students accordingly.  Statisticians use numerical methods to estimate derivatives and maximize likelihood; why should we not teach logistic regression early just because we can't use calculus to derive closed-form formulas for the parameter estimates?

The imperatives so far have focused on the abstract deductive structure of statistics. Part of what energizes our subject is the irresolvable tension between that structure and its application in context.

### 5.4  Exploit context

Statistics and its teaching have a history of success dealing with uncertainty.  We can quantify it, model it, and use it to assess inferences.  To date we have been much less successful dealing with ambiguity.  Ambiguity challenges our attempts to quantify it, we have no models for it, and there are few guides to help assess the ways it limits the scope of our inferences.

Ambiguity lives in the frontier territory where our methods and models meet context.  In abstract mathematics, context obscures structure; in applied data analysis context provides meaning.  This inherent tension between abstraction and meaning-in-context energizes our subject.  Applied statisticians stretch beyond existing theory to address new contexts; mathematical statisticians broaden their theories to bring new applied methodologies into the fold.

I see B&K's "Encourage deep interdisciplinary knowledge" (p. 108) as an exhortation to teach our students to experience, to appreciate, and to find creative opportunities in the energizing tension between abstraction and context. The remainder of this sub-section follows a progression from (a) standard uses of context, through (b) a more explicit focus on abstraction-as-process, to (c) a radical use of context as a vehicle for teaching abstract methods. This last is but a small step away from the final sub-section, Teach Through Research.

(a) *Three standard uses of context: interpretation, motivation, and direction*
- Interpretation. After computers began to make it possible to teach with real data, the better teachers and textbooks began asking students to "Tell what your results mean in the language of the applied context." This has become standard, although the depth and subtlety of the interpretation varies.
- Motivation: Use context to tell a story. For example, consider the details of the nasty academic food fight[36] between Pulitzer winner Jared Diamond and an opposing pair of ecologists. Their vitriolic exchanges are more salacious gossip than statistical pedagogy, but taking time to present excerpts can set the stage for an unfolding drama in which hypothesis testing and MCMC are major players.
- Direction. The *Statistical Sleuth* (Ramsey and Shafer 1996) provides survival data, by age and sex, for members of the Donner Party and asks whether, after adjusting for age, men were more likely than women to survive the rigors of their winter crossing of the Sierra Nevada. One can present the data and question on the first day of a course in statistical theory and spend the entire semester developing the answer, starting with estimation for a single binomial and ending with iterative fitting of the logistic model and asymptotic properties of the parameter estimates. Having an ongoing unanswered applied question not only provides motivation, but also continuity.

*(b) Abstraction as process.*
Plato's sense that mathematics is easy and theory of government is hard can be explained by the challenge of modeling: How can we penetrate the thicket of context to get to the trunk and branches of structure? When I first started teaching design of experiments, I was surprised by how hard students found it to identify the structure of an experiment from a verbal description: What was the response? What were the treatment factors? Were there blocks? How many sizes of experimental units were there? What would a rectangular format for the data look like? Helping students to learn how to parse the verbal description gave me a deep appreciation for how important it is to help students learn to *recognize* abstract structure. Our *teaching* tends to follow a top-down mathematical paradigm of abstraction first, illustration after. Context reminds us that *knowledge* tends to be built from the bottom up. Our curriculum needs to systematize the teaching of abstraction-as-process.

*(c) A radical use of context as primary vehicle.*
At Wesleyan University psychologist Lisa Dierker and her colleagues teach an unusual first statistics course to students in the natural and social sciences, students who need to learn statistics but have chosen not to take a more standard course offered by the mathematics department (Dierker *et al.*, 2013). These students tend to be those for whom a mathematically based exposition is a barrier to understanding. Nevertheless, by half-way through the semester these students are learning about multiple linear and logistic regression along with other methods. (Section 5.5e gives more detail.)

Skeptics may ask, "What are they actually *learning*?" and surely that question is legitimate. At the same time we should be careful not to prejudge the answer. As teachers of statistics we have mostly come from a tradition of mathematically-based, top-down exposition. B&K are clear about the shortcomings of the established approach, I find them persuasive, and so I suggest we need more innovation in the spirit of the Wesleyan course, and more data about alternative approaches.

The various goals I have suggested for context come together in the final imperative.

### 5.5  Teach Through Research

This imperative comes last because I consider it most important among the five, the one served by the other four, the culminating principle that connects our curriculum to its roots in the Sciences. In the Humanities, students in a first course engage with original sources. You don't just *prepare* your students to read Austen; they *read* Austen. You don't just *prepare* students to hear Bach; they *hear* Bach. Our statistics curriculum should follow those examples: Our job is not to *prepare* students to use data to answer a question that matters; our job is to help them use data to *answer* a question that matters. In short, *teach through research*. This may be the biggest exogenous challenge to our profession, the least explored in our undergraduate curriculum, and the most promising for rethinking what we teach. Accordingly, I am particularly excited about the possibilities suggested by Dierker (2013) and Metz (xref) in (e) and (f) below.

The imperative to teach through research rests on two theses: (1) If a teacher can motivate a student to learn, the main job is done and the learning will follow. (2) Nothing motivates students like choosing their own question and being the first to offer an answer. Taking these as postulates, we should be driven to ask, "How low can we go?" More formally, how early in the curriculum can we teach through research? B&K have issued the challenge: "Minimize prerequisites to research … There has been a tendency in statistics to have students first understand, then do. But this sequence can be reversed … " (p. 108). As part of rethinking our curriculum, we need to take this challenge seriously. To what extent can we minimize prerequisites to research? How early in the curriculum *can* we (and how early *should* we) rely on research-based learning?

(a) *First year of graduate school?* Students arrive with considerable technical background, but delaying the research experience until graduate school won't address the recruiting concerns about supply and pipeline that B&K write about.

(b) *Senior year of college?* Many seniors majoring in the natural and social sciences do research projects and write theses. It would seem straightforward to offer a parallel statistics course in which students report to each other on their projects and work with each other on statistical aspects of their research. Such a course would help students headed for graduate school in our client disciplines, but would probably not recruit many additional students to graduate school in statistics.

(c) *Upper-division undergraduates?* In the CIR program at St. Olaf College (L+ p. 59) "Students learn about statistics by doing statistics in authentic collaborations with researchers from other disciplines …" (In a similar spirit, see Spurrier *et al.* 1999 and Spurrier 2000, 2001.) The St. Olaf program is open on a competitive basis to all students who have completed a course in applied regression (p. 62) and it has a remarkable record of sending students to graduate school in statistics, so I hope readers will want to read about the St. Olaf program and think about whether and how it can be modified to fit their local situation. Almost surely some modification will be required, because the CIR program has benefited from substantial grant support, from the overtime hours of many energetic faculty, and from an unusual campus culture that puts high value on the mathematical sciences,

(d) *Lower division undergraduate*? Kuiper and Sklar (2013) is a textbook for a project-based second statistics course using real problems and data sets from recent scientific research such as the controversy about the hockey stick pattern and global warming. The book is structured as a sequence of guided investigations based on a given collection of problems, which does not qualify as research in the sense of B&K, but may for that same reason be more attractive as a starting point for teachers attracted to the principle but not yet ready to take the full plunge.

(e) *First statistics course*? (Dierker 2013). In this radical variant of the standard introductory course, each student learns statistics through an individual research experience. At the beginning of the semester, students learn about a variety of very large, very rich data sets such as the General Social Survey. They are asked to choose one of the data sets based on their interest, and to formulate a conjecture. Next, they do a literature search and narrow their conjecture to formulate a research hypothesis that can be investigated using cases and variables extracted from their chosen data base. For the remainder of the semester they learn statistical methods, decide which might be useful, analyze their data, and present their results and conclusions at a poster session where they give short presentation to outside evaluators. This constitutes their final examination. Thus Dierker's course embeds statistical thinking within the scientific process, and provides students with archival data of genuine import, while at the same time giving them a hands-on sense of owning their research.

(f) *Middle school*? Kathleen Metz (xref) has studied the learning by sixth graders about randomness and the scientific method, based on their designing, conducting, and presenting their own research projects. For example, students have access to a large terrarium of crickets. Each research group formulates a testable hypothesis about cricket behavior, designs a study, collects data, summarizes the results, and presents a summary of their data and their conclusions at a poster session.

(g) *Elementary school*. I remember visiting an elementary school with my 6-year-old daughter. Third grade students were conducting a survey, going from class to class, asking a set of questions and recording the results. It was in no way research in the sense of B&K, but to those students it was. They were trying to be the first to find answers to real-world questions that they themselves had chosen.

As these examples make clear, there is a tradeoff among (1) how early students are given an opportunity to learn by doing, (2) how much background they will bring to the enterprise, and (3) how deep an understanding they will take away from it. The existence of a tradeoff is, in my opinion, yet another reason to rethink the whole curriculum. We need more experience, and more research, to find effective ways to make intelligent decisions about the tradeoff, keeping in mind that the research experience is for many students a powerful motivation to learn statistics.

## 6. MOVING AHEAD: A SUMMARY AND TWO CAVEATS

Table 2 offers a provisional summary of principles for thinking ahead, one which I hope readers will find useful, albeit useful only in the sense of one more enthusiastic oar in the ongoing regatta.

Mathematics:
- Is no longer needed as a computational engine
- Still serves as a (partial) unifying frame
- Still serves as a path to deep understanding
- Still leads imagination to discover new methods
- But still grips with a dead hand on the undergraduate curriculum
- And still bars early access to the experience of research

Computation:
- Replaces mathematics as numerical engine
- Offers a direct, intuitive path to traditional concepts
- Opens an easy way to explore large data sets in the creative spirit of Tukey
- Encourages empirical evaluation of methods, e.g., via cross validation
- Offers entrée to original discovery in the spirit of research

Context:
- Opens the door to research questions that matter
- Motivates students to want to learn
- Complements and supports technical understanding

Taking these principles as one glimpse at blue sky, I conclude with two grounding caveats.

### 6.1 All curriculum is local

Curriculum unavoidably involves decisions about scarce resources, so curricular innovation cannot escape being political, and of course "all politics is local" (O'Neill and Hymel, 1995). Curriculum is political for economic reasons because, averaged over the long term, faculty FTEs and course offerings are at best a zero-sum game.[37] Thus changing curriculum, like moving a graveyard, depends on local conditions: Whose cherished ancestry is uprooted by the change? *A la* Moore (see below), the obverse is also true: if you want something to stay there, you've got to get it planted deep down.

### 6.2 For change to endure, you have to institutionalize it.

To paraphrase David Moore (1999), "All innovations succeed in the hands of the innovator. For real change to succeed, you have to institutionalize it." What can we do to institutionalize reform? Harder yet, how can we ensure that change in good directions will endure before we have a clear sense of which directions are the good ones? Here are just two sets of thoughts, under the headings publications and ASA.

- *Publications.*
  *The American Statistician* might begin a new and regular sub-section of the Teacher's Corner devoted to new courses, with short summaries in the printed journal, and on-line supplements and a forum for on-line comments. Other journals could do something similar (JSE, TISE, SERJ). For example, SERJ could publish course descriptions, together with a call for on-line comments regarding assessment and research. One goal of having a regular sub-section devoted to new courses would be to publicize innovations. A second goal would be to provide an incentive for innovation, and a way to help authors get scholarly credit for curricular work.

- *ASA.* Although I begin here with a complaint about ASA and four-year colleges, I am optimistic that we are beyond that now, and optimistic about our future.
  - o Four-year colleges. I find B&K, the Horton report, and this special issue of *TAS* especially encouraging in light of my 35 years teaching statistics to undergraduates, during which time I heard repeatedly from colleagues who taught in graduate programs some variant of "Just send us students with strong mathematics backgrounds; we'll teach them statistics." Is it any wonder we have had a pipeline problem: If you shut down the pump, where do you expect the water to come from?

    As an organization ASA has in the past sometimes fallen short in this area by excluding four-year college statisticians from its activities related to education. For example, the only publication about statistics education cited by B&K is the report *Modern Interdisciplinary <u>University</u> Statistics Education* (Kettenring 1994, my emphasis). The panel that produced this report included *no* statisticians from four-year colleges. A decade later ASA formed a task force on education with *no* representation from four-year colleges. I served on the Board at the time, and suggested that to the extent that the Association was concerned about supply, it might be good to include someone from an undergraduate institution. The task force membership did not change. More positively, however, the Board did at another time approve funding for a strategic initiative proposed by college statistician Thomas Moore to bring together educators from both colleges and universities.

    In this context it is especially bracing to see in B&K what I read as an implicit exhortation for our profession to pay more attention to undergraduate education. Until we statisticians join all the other sciences in supporting undergraduate teaching of our subject, we deprive ourselves of a vital source of future colleagues. Our membership will continue to be numerically challenged while potential undergraduate statistics majors choose instead to major in a subject that offers them a chance to learn through research.
  - o The Board of Directors might consider a statement to funding agencies and journals, citing the Horton report and stressing the particular importance of supporting and publicizing curricular experimentation outside the mainstream of undergraduate statistics education at this time of rapid change in the development of our subject.
  - o The Board might also consider a follow-up task force, this time specifically on articulation, experimentation, and implementation at the undergraduate level.

In conclusion, as we aim for a new synthesis, we should embrace a range of goals. Short term, we can work to implement the recommendations of the Horton report. Middle term, we can all think about ways to flatten prerequisites, seek depth, embrace computation, exploit context, and teach through research. Longer term, we can all experiment and evaluate, question everything, and take nothing for granted. You don't have to be a Martin Luther. You can always find a picnic that begs for a skunk.

**REFERENCES**

Notes:
(1) The dates in the list below correspond to publication dates, even though in many instances the original version of the work came much earlier. A canonical example is Bernoulli's work in 1692, published posthumously in 1713. More recently, I know first hand that Tukey (1977) is the published version of a book that originally circulated in mimeographed form many years earlier, and that Nemenyi and Dixon (1977) is based on a mimeographed book that Peter Nemenyi wrote and taught from during the 1960s. Almost surely, something similar is true of many of the other innovative textbooks cited here, but the reach of history exceeds the grasp of my first-hand knowledge. Where innovation is the context, I have tried to find and cite the earliest published version; where current teaching is the context, I have cited the most recent edition.

(2) The list below, though extensive, is not meant to be exhaustive. I am not a statistics historian, and my approach in this essay is subjective and hortatory. I use my sense of the past for argument, more like a lawyer than an archeologist. All the same, I hope the bibliography will be a useful source for others who care about statistics education.

---

[1] Through Ernst 2004, Nemenyi and Dixon 1977, Mosteller and Rourke 1973, and Noether 1971, all the way back to Fisher 1936 and Pitman 1937.

[2] Although I think the St. Olaf curriculum described in L+ p. 60 offers a new and effective model for making use of a prerequisite structure.

[3] Most versions of our upper-division course in probability (a mathematics course) still belong in the curriculum, there is still considerable value to be found in some elements of our 60-year-old course in mathematical statistics (but much needs to be changed), and an introductory course along the lines of the Advanced Placement syllabus (College Board 2010) can still be worthwhile as one possible starting place.  These courses remain serviceable.

[4] To paraphrase Bernoulli, how many trials are required to reach "moral certainty" about the value of an unknown probability? (Hacking, 1975)  For Bernoulli's derivation, in modern notation, see Uspensky, 1937, pp. 96 – 101.  Decades after Bernoulli, de Moivre (1733) improved on his solution, deriving a normal approximation to the distribution of $p - \hat{p}$ for the special case $p = \frac{1}{2}$.  After that, Laplace generalized de Moivre's result to all values of $p$.  (See Adams 2009)  Here, also, mathematics was a computational engine.

[5] In his mathematical vision, Fisher had precursors (Laplace's use of Bayesian methods, Galton's and Pearson's parametric families), but those were hardly determinative of mathematics as a unifying principle.  Fisher's pioneering use of mathematics as unifier was followed immediately and vigorously by a trio who embraced the principle but challenged its implementation in Fisher's hands.  Neyman, Pearson the younger, and Wald countered with a unifying theory based on operating characteristics – the long-run behavior of decision rules for hypothesis testing and interval estimates.  The two sides never reconciled, but all the same, together they established mathematical thinking as a unifying principle for statistical inference.

[6]    By halfway through the 20th century, statistics was recognized as important in a practical sense, especially at land grant colleges, thanks to Fisher, Snedecor, Cochran and (Gertrude) Cox.  At the same time, thanks to Fisher, Neyman, Pearson, Wald, and others, statistics was gaining recognition as an important area of applied mathematics.  Even so, the teaching of statistics was narrowly blinkered:  applied courses were taught at land grant colleges, and theoretical courses were taught in mathematics departments.  To exaggerate, statistics was still seen, down the nose, as a pursuit either for intellectually ambitious farm scientists or for slumming mathematicians.  Only at mid-century did the camel of statistics begin to stick its nose into the undergraduate tent of curricular legitimacy.  That overdue intrusion was led by mathematicians John Kemeny and Laurie Snell, at Dartmouth via Princeton, by Fred Mosteller at Harvard, and others -- including statisticians Paul Hoel (1947) at UCLA, Samuel Wilks (1948 at Princeton), Jerzey Neyman (1950) at UC Berkeley, D. A. S. Fraser (1958) at the University of Toronto, and Robert Hogg (1959) at the University of Iowa -- introduced mathematical statistics into the curriculum.[6] Kemeny and Snell (1959) established probability as a new part of the lower division undergraduate mathematics curriculum with their book *Introduction to Finite Mathematics*. At roughly the same time, Fred Mosteller began teaching a course based on materials that were to become *Probability with Statistical Applications* (1961).  By putting the emphasis on probability, and relegating statistics to the status of applications, Mosteller brought legitimacy to teaching statistical ideas as part of a mathematics curriculum.

[7] Granted, the "dark satanic mills" of the industrial revolution, and later the railroad, both had major impacts, but not intellectually, and not on the scale of the printing press.

[8] N&TL (p. 99) quote John Hartigan saying in 2005 that the internet and the S language are the two greatest contributions to statistics over the last 40 years."

[9] In a sense, context and mathematics provide complementary ways to introduce and explain statistics, and reliance on mathematics as a source of respectability has led many teachers and authors to underestimate the importance of context.

[10] For example, Fisher 1925, 1935 and Snedecor 1937.

[11] Acton 1959, Bliss 1967, 1970, Daniel and Wood 1971, Colton 1974, Tufte 1974.

[12] e.g., Moroney 1952, Bross 1953, Wallis and Roberts 1956, and later Federer (1972) and Tanur *et al.* 1972.

[13] Freedman *et al.* (1978) was an early standout, but there are now many to choose from.

[14] Larsen and Stroup 1976, Chatterjee and Price 1977.

[15] Kuiper and Sklar 2013, Spurrier *et al.* 1999.

[16] Astronomy:  the orbits of the moon and of Saturn and Jupiter.  Geodesy:  The shape of the earth.

[17] Even the nonparametric randomization-based t-test requires justification for taking all permutations as equally likely.

[18] Imagine yourself trying to teach each example in an introductory course on the first day of class.  The t-test relies on so many prerequisite concepts (randomness, normal distribution, mean, standard deviation, standard error of the difference of two means, …) that it typically comes late in the semester.  The idea of a classification tree is intuitive and easy to explain.  A teacher could hand out a data set, quickly explain the goal of the analysis, and leave the remainder of the class period for students to create and evaluate their own algorithms.  (See Wagaman 2013.)

[19] For example, Tukey's "break tables," however, are now largely lost to memory. (These tables allowed students to transform data quickly and by hand, to roots, logs, and reciprocals.)  My point in citing break tables is this:  Tukey knew that effective exploration often required messy computations, such as converting to logs.  His ingenuity led to a paper/pencil end run.  But imagine what he might have done if technology had not been a barrier?

[20] One important feature of CEDA missed by this parallel is that whereas Tukey's EDA is typically agnostic about where and how to explore, "algorithmic thinking" in Breiman's sense is often very sharply focused on a known goal, e.g., to separate "spam" from other e-mail.

[21] B&K addresses a broad range of important issues. In this section I focus narrowly on the apparent difference between that article and Breiman's, but in Section 5 I return to many other points from B&K.

[22] Few if any other branches of mathematics had to wait three centuries for a formal axiomatic formulation.

[23] Indeed, their only cited reference to a publication about statistics education (Kettenring, 1994) is noteworthy for the absence of any mention of statistics at four-year colleges.)

[24] I want to preempt a possible misunderstanding. In arguing that we flatten prerequisites, and that our curriculum is too-often shaped by too-great a dependence on underlying mathematical structures, I do not argue that we should *avoid* mathematics, only that we should think carefully about balance. At base *the issue is not mathematics but gratuitous linearity*. Statisticians also thrill to the glorious structures; and mathematicians also worry about the harmful effects of prerequisites; see O'Shea and Pollatsek 1997.

[25] Some argue, with justification, that the five should in fact be six, inserting an introductory applied course somewhere in the sequence. Either way, few students ever become eligible to take mathematical statistics.

[26] In statistics courses, it often happens that we need partial derivatives. A likelihood function has two independent variables; so does the sum of squares for simple linear regression, regarded as a function of the slope and intercept. We might insist that students take a course in multivariable calculus, but it if they have seen a derivative, it is not hard to take five or ten minutes in class to talk about partial derivatives, their use, and possible complexities.

[27] Kaplan's course neither requires nor teaches the geometry suggested by the phrase "Euclidean n-space." Instead, Kaplan exploits the geometry as a schematic way to represent variables as vectors and in that way to think about how they are related in a regression model.

[28] Moore and McCabe 1984, Samuels and Witmer 2011.

[29] Kirk 1994, Sokal and Rohlf 2011.

[30] Kleinbaum *et al*. 1988 and Fox 2008.

[31] Of course I have ignored many subtleties, e.g., elementary outcomes that are not equally likely, and infinite sample spaces, but these can be introduced as concepts, and then simulated, to arbitrary precision, using finite spaces of equally likely outcomes.

[32] See Hald 1998 for a more exact translation of what Laplace actually wrote.

[33] If $T$ is unbiased then $\text{Corr}^2(T, \text{Score}) \leq 1 \Rightarrow \text{Var}(T) \geq 1/\text{Var}(\text{Score})$.

[34] One factor is inside a second if each group of the first factor is a subset of some group of the second factor.

[35] It lends itself to physical simulation with cups of marbles -- each cup is a value of $\theta$, and the color of a randomly chosen marble is the observation $y$.

[36] See Cobb and Chen 2002.

[37] A new FTE in statistics must be offset by a lost FTE in music history or Renaissance literature or Women's Studies, and must compete for the slot with bioinformatics or computer science or economics. Even more locally, within a Liberal Arts college department of mathematics and statistics, a new course in applied regression or Markov chain Monte Carlo competes with courses in number theory or differential geometry. Moreover, if the department statistician teaches a section of required calculus instead of an applied elective in correlated data, that frees up a mathematician to teach a course in Lie groups or Galois theory.